# PARIS: Probabilistic Alignment of Relations, Instances, and Schema

Fabian M. Suchanek
INRIA Saclay – Île-de-France
4 rue Jacques Monod
91893 Orsay Cedex, France
fabian@suchanek.name

Serge Abiteboul
INRIA Saclay – Île-de-France
4 rue Jacques Monod
91893 Orsay Cedex, France
serge.abiteboul@inria.fr

Pierre Senellart
Institut Télécom;
Télécom ParisTech; CNRS LTCI
75634 Paris Cedex 13, France
pierre.senellart@telecom-paristech.fr

## ABSTRACT

One of the main challenges that the Semantic Web faces is the integration of a growing number of independently designed ontologies. In this work, we present PARIS, an approach for the automatic alignment of ontologies. PARIS aligns not only instances, but also relations and classes. Alignments at the instance level cross-fertilize with alignments at the schema level. Thereby, our system provides a truly holistic solution to the problem of ontology alignment. The heart of the approach is probabilistic, i.e., we measure degrees of matchings based on probability estimates. This allows PARIS to run without any parameter tuning. We demonstrate the efficiency of the algorithm and its precision through extensive experiments. In particular, we obtain a precision of around 90 % in experiments with some of the world's largest ontologies.

## 1. INTRODUCTION

*Motivation.* An ontology is a formal collection of world knowledge. In this paper, we use the word *ontology* in a very general sense, to mean both the schema (classes and relations), and the instances with their assertions. In recent years, the success of Wikipedia and algorithmic advances in information extraction have facilitated the automated construction of large general-purpose ontologies. Notable endeavors of this kind include DBPEDIA [2], KNOWITALL [10], WIKITAXONOMY [26], and YAGO [30], as well as commercial services such as freebase.com, trueknowledge.com, and wolframalpha.com. These ontologies are accompanied by a growing number of knowledge bases[1] in a wide variety of domains including: music[2], movies[3], geographical data[4], publications[5], medical and biological data[6], or government data[7].

Many of these ontologies contain complementing data. For instance, a general ontology may know who discovered a certain enzyme, whereas a biological database may know its function and properties. However, since the ontologies generally use different terms (identifiers) for an entity, their information cannot be easily brought together. In this respect, the ontologies by themselves can be seen as isolated islands of knowledge. The goal of the Semantic Web vision is to interlink them, thereby creating one large body of universal ontological knowledge [5, 6]. This goal may be seen as a much scaled-up version of record linking, with challenges coming from different dimensions: (i) unlike in record linkage, both instances and schemas should be reconciled; (ii) the semantics of the ontologies have to be respected; (iii) the ontologies are typically quite large and complex. Moreover, we are interested in performing the alignment in a fully automatic manner, and avoid tedious tuning or parameter settings.

A number of recent research have investigated this problem. There have been many works on entity resolution, i.e., on what is traditionally known as the "A-Box" [1, 4, 12, 17, 18, 25, 27, 28, 31]. In another direction, much research has focused on schema alignment, i.e., on the so-called "T-Box" [3, 14, 20, 21, 34]. However, in recent years, the landscape of ontologies has changed dramatically. Today's ontologies often contain both a rich schema and, at the same time, a huge number of instances, with dozens of millions of assertions about them. To fully harvest the mine of knowledge they provide, their alignment has to be built on cross-fertilizing the alignments of both instances and schemas.

In this paper, we propose a new, holistic algorithm for aligning ontologies. Our approach links not just related entity or relationship instances, but also related classes and relations, thereby capturing the fruitful interplay between schema and instance matching. Our final aim is to discover and link identical entities automatically across ontologies on a large scale, thus allowing ontologies to truly complement each other.

---

[1] http://www.w3.org/wiki/DataSetRDFDumps
[2] http://musicbrainz.org/
[3] http://www.imdb.com/
[4] http://www.geonames.org/

[5] http://www.informatik.uni-trier.de/~ley/db
[6] http://www.uniprot.org/
[7] http://www.govtrack.us/, http://source.data.gov.uk/data/

*Contribution.* The contribution of the present paper is three-fold:

1. We present PARIS[8], a probabilistic algorithm for aligning instances, classes, and relations simultaneously across ontologies.

2. We show how this algorithm can be implemented efficiently and that it does not require any tuning

3. We prove the validity of our approach through experiments on real-world ontologies.

The paper is organized as follows. Section 2 provides an overview of related work. We then introduce some preliminaries in Section 3. Section 4 describes our probabilistic algorithm and Section 5 its implementation. Section 6 discusses experiments. To ease the reading, some technical discussions are postponed to the appendix.

## 2. RELATED WORK

*Overview.* The problem of ontology matching has its roots in the problem of identifying duplicate entities, which is also known as record linkage, duplicate detection, or co-reference resolution. This problem has been extensively studied in both database and natural language processing areas [7, 9]. These approaches are less applicable in the context of ontologies for two reasons. First, they do not consider the formal semantics that ontologies have (such as the *subclassOf* taxonomy). Second, they focus on the alignment of instances and do not deal with the alignment of relations and classes.

There are a number of surveys and analyses that shed light on the problem of record linking in ontologies. Halpin et al. [15] provide a good overview of the problem in general. They also study difficulties of existing *sameAs*-links. These links are further analyzed by Ding et al. [8]. Glaser, Jaffri, and Millard [13] propose a framework for the management of co-reference in the Semantic Web. Hu et al. [19] provide a study on how matches look in general.

*Schema Alignment.* Traditional approaches to ontology matching have focused mostly either on aligning the classes (the "T-Box") or on matching instances (the "A-Box"). The approaches that align the classes are manifold, using techniques such as sense clustering [14], lexical and structural characteristics [21], or composite approaches [3]. Unlike PARIS, these approaches can only align classes and do not consider the alignment of relations and instances.

Most similar to our approach in this field are [20] and [34], which derive class similarity from the similarities of the instances. Both approaches consider only the equivalence of classes and do not compute subclasses, as does PARIS. Furthermore, neither can align relations or instances.

*Instance Matching.* There are numerous approaches to match instances of one ontology to instances of another ontology. Ferrara, Lorusso, and Montanelli [12] introduce this problem from a philosophical point of view. Different techniques are being used, such as exploiting the terminological structure [25], logical deduction [27], declarative languages [1], relational clustering [4], or a combination of logical and numerical methods [28]. The SIG.MA engine [31] uses heuristics to match instances. Perhaps closest to our approach is [17], which introduces the concept of functionality. Different from their approach, PARIS does not require an additional smoothening factor.

---
[8]Probabilistic Alignment of Relations, Instances, and Schema

The SILK framework [33] allows specifying manual mapping rules. The OBJECTCOREF approach by Hu, Chen, and Qu [18] allows learning a mapping between the instances from training data. With PARIS, we aim at an approach that uses neither manual input nor training data. We compare some of the results of OBJECTCOREF to that of PARIS on the datasets of the ontology alignment evaluation initiative [11] in Section 6. Hogan [16] matches instances and proposes to use these instances to compute the similarity between classes, but provides no experiments. Thus, none of these approaches can align classes and relations like PARIS.

*Holistic Approaches.* Only very few approaches address the cause of aligning both schema and instances: the RIMOM [22] and ILIADS [32] systems. Both of these have only been tested on small ontologies. The RIMOM system can align classes, but it cannot find *subclassOf* relationships. Furthermore, the approach provides a bundle of heuristics and strategies to choose from, while PARIS is monolithic. None of the ontologies the ILIADS system has been tested on contained full-fledged instances with properties. In contrast, PARIS is shown to perform well even on large-scale real-world ontologies with millions of instances.

## 3. PRELIMINARIES

In this section, we recall the notions of ontology and of equivalence. Finally, we introduce the notion of functionality as one of the key concepts for ontology alignment.

*Ontologies.* We are concerned with ontologies available in the Resource Description Framework Schema (RDFS [35]), the W3C standard for knowledge representation. An RDFS ontology builds on *resources*. A resource is an identifier for a real-world object, such as a city, a person, or a university, but also the concept of mathematics. For example, *London* is a resource that represents the city of London. A *literal* is a string, date or number. A *property* (or *relation*) is a binary predicate that holds between two resources or between a resource and a literal. For example, the property *isLocatedIn* holds between the resources *London* and *UK*. In the RDFS model, it is assumed that there exists a fixed global set $\mathcal{R}$ of resources, a fixed global set $\mathcal{L}$ of literals, and a fixed global set $\mathcal{P}$ of properties. Each resource is described by a URI. An RDFS *ontology* can be seen as a set of triples $O \subset \mathcal{R} \times \mathcal{P} \times (\mathcal{R} \cup \mathcal{L})$, called *statements*. In the following, we assume given an ontology $O$. To say that $\langle x, r, y \rangle \in O$, we will write $r(x, y)$ and we call $x$ and $y$ the *arguments* of $r$. Intuitively, such a statement means that the relation $r$ holds between the entities $x$ and $y$. We say that $x, y$ is a *pair* of $r$. A relation $r^{-1}$ is called the *inverse* of a relation $r$ if $\forall x, y : r(x, y) \Leftrightarrow r^{-1}(y, x)$. We assume that the ontology contains all inverse relations and their corresponding statements. Note that this results in allowing the first argument of a statement to be a literal, a minor digression from the standard.

An RDFS ontology distinguishes between classes and instances. A class is a resource that represents a set of objects, such as, e.g., the class of all singers, the class of all cities or the class of all books. A resource that is a member of a class is called an *instance* of that class. We assume that the ontology partitions the resources into classes and instances.[9] The *rdf:type* relation connects an instance to a class. For example, we can say that the resource *Elvis* is a member of the class of singers: *rdf:type(Elvis, singer)*.

A more specific class $c$ can be specified as a *subclass* of a more general class $d$ using the statement *rdfs:subclassOf(c,d)*. This means that, by inference, all instances of $c$ are also instances of $d$. Likewise, a relation $r$ can be made a sub-relation of a relation $s$ by the statement *rdfs:subpropertyOf(r,s)*. This means that, by inference again, $\forall x, y : r(x, y) \Rightarrow s(x, y)$. We assume that all such inferences have been established and that the ontologies are available in their *deductive closure*, i.e., all statements implied by the subclass and sub-property statements have been added to the ontology.

*Equivalence.* In RDFS, the sets $\mathcal{P}$, $\mathcal{R}$, and $\mathcal{L}$ are global. That means that some resources, literals, and relations may be *identical* across different ontologies. For example, two ontologies may contain the resource *London*, therefore share that resource. (In practice, *London* is a URI, which makes it easy for two ontologies to use exactly the same identifier.) The semantics of RDFS enforces that these two occurrences of the identifier refer to the same real-world object (the city of London). The same applies to relations or literals that are shared across ontologies. Conversely, two different resources can refer to the same real-world object. For example, *London* and *Londres* can both refer to the city of London. Such resources are called *equivalent*. We write $Londres \equiv London$.

The same observation applies not just to instances, but also to classes and relations. Two ontologies can talk about an identical class or relation. They can also use different resources, but refer to the very same real-world concepts. For example, one ontology can use the relation *wasBornIn* whereas another ontology can use the relation *birthPlace*. An important goal of our approach is to find out that $wasBornIn \equiv birthPlace$.

In this paper, we make the following assumption: *a given ontology does not contain equivalent resources*. That is, if an ontology contains two instances $x$ and $x'$, then we assume $x \not\equiv x'$. We assume the same for relations and classes. This is a reasonable assumption, because most ontologies are either manually designed [23, 24], or generated from a database (such as the datasets mentioned in the introduction), or designed with avoiding equivalent resources in mind [30]. If the ontology does contain equivalent resources, then our approach will still work. It will just not discover the equivalent resources within one ontology. Note that, under this assumption, there can never be a chain of equivalent entities. Therefore, we do not have to take into account transitivity of equivalence.

*Functions.* A relation $r$ is a *function* if, for a given first argument, there is only one second argument. For example, the relation *wasBornIn* is a function, because one person is born in exactly one place. A relation is an *inverse function* if its inverse is a function. If $r$ is a function and if $r(x, y)$ in one ontology and $r(x, y')$ in another ontology, then $y$ and $y'$ must be equivalent. In the example: If a person is born in both *Londres* and *London*, then $Londres \equiv London$. The same observation holds for two first arguments of inverse functions.

---

[9]RDFS allows classes to be instances of other classes, but in practice, this case is rare.

As we shall see, functions play an essential role in deriving alignments between ontologies. Nevertheless, it turns out that the precise notion of function is too strict for our setting. This is due to two reasons:

- First, a relation $r$ ceases to be a function as soon as there is one $x$ with $y$ and $y'$ such that $r(x, y)$ and $r(x, y')$. This means that just one erroneous fact can make a relation $r$ a non-function. Since real-world ontologies usually contain erroneous facts, the strict notion of function is not well-suited.

- Second, even if a relation is not a function, it may contribute evidence that two entities are the same. For example, the relation *livesIn* is not a function, because some people may live in several places. However, a wide majority of people live in one place, or in very few places. So, if most people who live in *London* also live in *Londres*, this provides a strong evidence for the unification of *London* and *Londres*.

Thus, to derive alignments, we want to deal with "quasi-functions". This motivates introducing the concept of *functionality*, as in [17]. The *local functionality* of a relation $r$ for a first argument $x$ is defined as:

$$fun(r, x) = \frac{1}{\#y : r(x, y)} \quad (1)$$

where we write "$\#y : \varphi(y)$" to mean "$|\{y \mid \varphi(y)\}|$". Consider for example the relationship *isCitizenOf*. For most first arguments, the functionality will be 1, because most people are citizens of exactly one country. However, for people who have multiple nationalities, the functionality may be $\frac{1}{2}$ or even smaller. The *local inverse functionality* is defined analogously as $fun^{-1}(r, x) = fun(r^{-1}, x)$. Deviating from [17], we define the *global functionality* of a relation $r$ as the harmonic mean of the local functionalities, which boils down to

$$fun(r) = \frac{\#x : \exists y : r(x, y)}{\#x, y : r(x, y)} \quad (2)$$

We discuss design alternatives for this definition and the rationale of our choice in Appendix A. The *global inverse functionality* is defined analogously as $fun^{-1}(r) = fun(r^{-1})$.

## 4. PROBABILISTIC MODEL

### 4.1 Equivalence of Instances

We want to model the probability $\Pr(x \equiv x')$ that one instance $x$ in one ontology is equivalent to another instance $x'$ in another ontology. Let us assume that both ontologies share a relation $r$. Following our argument in Section 3, we want the probability $\Pr(x \equiv x')$ to be large if $r$ is highly inverse functional, and if there are $y \equiv y'$ with $r(x, y), r(x', y')$ (if, say, $x$ and $x'$ share an e-mail address). This can be written pseudo-formally as:

$$\exists r, y, y' : r(x, y) \land r(x', y') \land y \equiv y' \land fun^{-1}(r) \text{ is high} \quad (3)$$
$$\implies x \equiv x'$$

Using the formalization described in Appendix B, we transform this logical rule into a probability estimation for $x \equiv x'$

as follows:

$$\Pr_1(x \equiv x') := 1 - \prod_{\substack{r(x,y) \\ r(x',y')}} \left(1 - fun^{-1}(r) \times \Pr(y \equiv y')\right) \quad (4)$$

In other words, as soon as there is one relation $r$ with $fun^{-1}(r) = 1$ and with $r(x,y)$, $r(x',y')$, and $\Pr(y \equiv y') = 1$, it follows that $\Pr_1(x \equiv x') = 1$. We discuss a design alternative in Appendix C.

Note that the probability of $x \equiv x'$ depends recursively on the probabilities of other equivalences. These other equivalences may hold either between instances or between literals. We discuss the probability of equivalence between two literals in Section 5. Obviously, we set $\Pr(x \equiv x) := 1$ for all literals and instances $x$.

Equation (4) considers only positive evidence for an equality. To consider also evidence against an equality, we can use the following modification. We want the probability $\Pr(x \equiv x')$ to be small, if there is a highly functional relation $r$ with $r(x,y)$ and if $y \not\equiv y'$ for all $y'$ with $r(x',y')$. Pseudo-formally, this can be written as

$$\exists r, y : r(x,y) \land (\forall y' : r(x',y') \Rightarrow y \not\equiv y') \land fun(r) \text{ is high} \quad (5)$$
$$\implies x \not\equiv x.$$

This can be modeled as

$$\Pr_2(x \equiv x') := \prod_{r(x,y)} \left(1 - fun(r) \prod_{r(x',y')} \left(1 - \Pr(y \equiv y')\right)\right) \quad (6)$$

As soon as there is one relation $r$ with $fun(r) = 1$ and with $r(x,y)$, $r(x',y')$, and $\Pr(y \equiv y') = 0$, it follows that $\Pr_2(x \equiv x') = 0$. We combine these two desiderata by multiplying the two probability estimates:

$$\Pr_3(x \equiv x') := \Pr_1(x \equiv x') \times \Pr_2(x \equiv x') \quad (7)$$

In the experiments, we found that Equation (4) suffices in practice. However, we discuss scenarios where Equation (7) can be useful in Section 6.

### 4.2 Subrelations

The formulas we have just established estimate the equivalence between two entities that reside in two different ontologies, if there is a relation $r$ that is common to the ontologies. It is also a goal to discover whether a relation $r$ of one ontology is equivalent to a relation $r'$ of another ontology. More generally, we would like to find out whether $r$ is a sub-relation of $r'$, written $r \subseteq r'$.

Intuitively, the probability $\Pr(r \subseteq r')$ is proportional to the number of pairs in $r$ that are also pairs in $r'$:

$$\Pr(r \subseteq r') := \frac{\#x,y : r(x,y) \land r'(x,y)}{\#x,y : r(x,y)} \quad (8)$$

The numerator should take into account the resources that have already been matched across the ontologies. Therefore, the numerator is more appropriately phrased as:

$$\#x,y : r(x,y) \land \left(\exists x',y' : x \equiv x' \land y \equiv y' \land r'(x',y')\right) \quad (9)$$

Using again our formalization from Appendix B, this can be modeled as:

$$\sum_{r(x,y)} \left(1 - \prod_{r'(x',y')} \left(1 - (\Pr(x \equiv x') \times \Pr(y \equiv y'))\right)\right) \quad (10)$$

In the denominator, we want to normalize by the number of pairs in $r$ that have a counterpart in the other ontology. This is

$$\sum_{r(x,y)} \left(1 - \prod_{x',y'} \left(1 - (\Pr(x \equiv x') \times \Pr(y \equiv y'))\right)\right) \quad (11)$$

Thus, we estimate the final probability $\Pr(r \subseteq r')$ as:

$$\frac{\sum_{r(x,y)} \left(1 - \prod_{r'(x',y')} \left(1 - (\Pr(x \equiv x') \times \Pr(y \equiv y'))\right)\right)}{\sum_{r(x,y)} \left(1 - \prod_{x',y'} \left(1 - P(x \equiv x') \times \Pr(y \equiv y'))\right)\right)} \quad (12)$$

This probability depends on the probability that two instances (or literals) are equivalent.

One might be tempted to set $\Pr(r \subseteq r) := 1$ for all relations $r$. However, in practice, we observe cases where the first ontology uses $r$ where the second ontology omits it. Therefore, we compute $\Pr(r \subseteq r)$ as a contingent quantity.

We are now in a position to generalize Equation (4) to the case where the two ontologies do not share a common relation. For this, we need to replace every occurrence of $r(x',y')$ by $r'(x',y')$ and factor in the probabilities that $r' \subseteq r$ or $r \subseteq r'$. This gives the following value to be assigned to $\Pr(x \equiv x')$:

$$1 - \prod_{\substack{r(x,y) \\ r'(x',y')}} \left(1 - \Pr(r' \subseteq r) \times fun^{-1}(r) \times \Pr(y \equiv y')\right) \quad (13)$$
$$\times \left(1 - \Pr(r \subseteq r') \times fun^{-1}(r') \times \Pr(y \equiv y')\right)$$

If we want to consider also negative evidence as in Equation (7), we get for $\Pr(x \equiv x')$:

$$\left(1 - \prod_{\substack{r(x,y) \\ r'(x',y')}} \left(1 - P(r' \subseteq r) \times fun^{-1}(r) \times \Pr(y \equiv y')\right)\right.$$
$$\left. \times \left(1 - \Pr(r \subseteq r') \times fun^{-1}(r') \times \Pr(y \equiv y')\right)\right)$$
$$\times \prod_{\substack{r(x,y) \\ r'}} \left(1 - fun(r) \times \Pr(r' \subseteq r) \times \prod_{r'(x',y')} \left(1 - \Pr(x \equiv x')\right)\right)$$
$$\times \left(1 - fun(r') \times \Pr(r \subseteq r') \times \prod_{r'(x',y')} \left(1 - \Pr(x \equiv x')\right)\right)$$
$$(14)$$

This formula looks asymmetric, because it considers only $\Pr(r' \subseteq r)$ and $fun(r)$ one the one hand, and $\Pr(r \subseteq r')$ and $fun(r')$ on the other hand (and not, for instance, $\Pr(r' \subseteq r)$ together with $fun(r')$). Yet, it is not asymmetric, because each instantiation of $r'$ will at some time also appear as an instantiation of $r$. It is justified to consider $\Pr(r' \subseteq r)$, because a large $\Pr(r' \subseteq r)$ implies that $r'(x,y) \Rightarrow r(x,y)$. This means that a large $\Pr(r' \subseteq r)$ implies that $fun(r) < fun(r')$ and $fun^{-1}(r) < fun^{-1}(r')$.

If there is no $x', y'$ with $r'(x',y')$, we set as usual the last factor of the formula to one, $\prod_{r'(x',y')}(1 - \Pr(x \equiv x')) := 1$. This decreases $\Pr(x \equiv x')$ in case one instance has relations that the other one does not have.

To each instance from the first ontology, our algorithm assigns multiple equivalent instances from the second ontology, each with a probability score. For each instance from the first ontology, we call the instance from the second ontology with the maximum score the *maximal assignment*. If there are multiple instances with the maximum score, we break ties arbitrarily, so that every instance has at most one maximal assignment.

## 4.3 Subclasses

A class corresponds to a set of entities. One could be tempted to treat classes just like instances and compute their equivalence. However, the class structure of one ontology may be more fine-grained than the class structure of the other ontology. Therefore, we aim to find out not whether one class $c$ of one ontology is equivalent to another class $c'$ of another ontology, but whether $c$ is a subclass of $c'$, $c \subseteq c'$. Intuitively, the probability $\Pr(c \subseteq c')$ shall be proportional to the number of instances of $c$ that are also instances of $c'$:

$$\Pr(c \subseteq c') = \frac{\# \ c \cap c'}{\#c} \quad (15)$$

Again, we estimate the expected number of instances that are in both classes as

$$\mathbb{E}(\# \ c \cap c') = \sum_{x:type(x,c)} \left(1 - \prod_{y:type(y,d)} (1 - P(x \equiv y))\right) \quad (16)$$

We divide this expected number by the total number of instances of $c$:

$$\Pr(c \subseteq c') = \frac{\sum_{x:type(x,c)} \left(1 - \prod_{y:type(y,d)} (1 - P(x \equiv y))\right)}{\#x : type(x,c)} \quad (17)$$

The fact that two resources are instances of the same class can reinforce our belief that the two resources are equivalent. Hence, it seems tempting to feed the subclass-relationship back into Equation (13). However, in practice, we found that the class information is of less use for the equivalence of instances. This may be because of different granularities in the class hierarchies. It might also be because some ontologies use classes to express certain properties (*MaleSingers*), whereas others use relations for the same purpose (*gender = male*). Therefore, we compute the class equivalences only *after* the instance equivalences have been computed.

## 5. IMPLEMENTATION

### 5.1 Iteration

Our algorithm takes as input two ontologies. As already mentioned, we assume that a single ontology does not contain duplicate (equivalent) entities. This corresponds to some form of a domain-restricted unique name assumption. Therefore, our algorithm considers only equivalence between entities from different ontologies.

Strictly speaking, the functionality of a relation (Equation (2)) depends recursively on the equivalence of instances. If, e.g., every citizen lives in two countries, then the functionality of *livesIn* is $\frac{1}{2}$. If our algorithm unifies the two countries, then the functionality of *livesIn* jumps to 1. However, since we assume that there are no equivalent entities within one ontology, we compute the functionalities of the relations within each ontology upfront.

We implemented a fixpoint computation for Equations (12) and (13). First, we compute the probabilities of equivalences of instances. Then, we compute the probabilities for subrelationships. These two steps are iterated until convergence. In a last step, the equivalences between classes are computed by Equation (17) from the final assignment. To bootstrap the algorithm in the very first step, we set $\Pr(r \subseteq r') = \theta$ for all pairs of relations $r, r'$ of different ontologies. We chose $\theta = 0.10$. The second round uses the computed values for $\Pr(r \subseteq r')$ and no longer $\theta$.

We have not yet succeeded in proving a theoretical condition under which the iteration of Equations (12) and (13) reaches a fixpoint. In practice, we iterate until the entity pairs under the maximal assignments change no more (which is what we call convergence). In our experiments, this state was always reached after a few iterations. We note that one could always enforce convergence of such iterations by introducing a progressively increasing dampening factor.

Our model changes the probabilities of two resources being equal – but never the probability that a certain statement holds. All statements in both ontologies remain valid. This is possible because an RDFS ontology cannot be made inconsistent by equating resources, but this would not be the case any more for richer ontology languages.

### 5.2 Optimization

The equivalence of instances (Equation (13)) can be computed in different ways. In the most naïve setting, the equivalence is computed for each pair of instances. This would result in a runtime of $O(n^2m)$, where $n$ is the number of instances and $m$ is the average number of statements in which an instance occurs (a typical value for $m$ is 20). This implementation took weeks to run one iteration. We overcame this difficulty as follows.

First, we optimize the computation of Equation (13). For each instance $x$ in the first ontology, we traverse all statements $r(x, y)$ in which this instance appears as first argument. (Remember that we assume that the ontology contains all inverse statements as well.) For each statement $r(x, y)$, we consider the second argument $y$, and all instances $y'$ that the second argument is known to be equal to ($\{y' : \Pr(y \equiv y') > 0\}$). For each of these equivalent instances $y'$, we consider again all statements $r(x', y')$ and update the equality of $x$ and $x'$. This results in a runtime of $O(nm^2e)$, where $e$ is the average number of equivalent instances per instance (typically around 10). Equations (12) and (17) are optimized in a similar fashion.

Generally speaking, our model distinguishes *true* equivalences ($\Pr(x \equiv x') > 0$) from *false* equivalences ($\Pr(x \equiv x') = 0$) and *unknown* equivalences ($\Pr(x \equiv x')$ not yet computed). Unknown quantities are simply omitted in the sums and products of the equations. Interestingly, most equations contain a probability $\Pr(x \equiv x')$ only in the form $\prod(1 - P(x \equiv x'))$. This means that the formula will evaluate to the same value if $\Pr(x \equiv x')$ is unknown or if $\Pr(x \equiv x') = 0$. Therefore, our algorithm does not need to store equivalences of value 0 at all.

Our implementation thresholds the probabilities and assumes every value below $\theta$ to be zero. This greatly reduces the number of equivalences that the algorithm needs to store. Furthermore, we limit the number of pairs that are evaluated in Equations (12) and (17) to 10,000. For each computation, our algorithm considers only the equalities of the previous maximal assignment and ignores all other equalities. This reduces the runtime by an order of magnitude without affecting much the relation inclusion assessment.

We stress that all these optimizations have for purpose to decrease the running time of the algorithm *without signif-*

*icantly affecting the outcome of the computation.* We have validated in our experiments that it is indeed the case.

Our implementation is in Java, using the Java Tools developed for [29] and Berkeley DB. We used the Jena framework to load and convert the ontologies. The algorithm turns out to be heavily IO-bound. Therefore, we used a solid-state drive (SSD) with high read bandwidth to store the ontologies. This brought the computation time down from the order of days to the order of hours on very large ontologies. We considered parallelizing the algorithm and running it on a cluster, but it turned out to be unnecessary.

### 5.3 Literal Equivalence

The probability that two literals are equal is known a priori and will not change. Therefore, such probabilities can be set upfront (*clamped*), for example as follows:

- The probability that two numeric values of the same dimension are equal can be a function of their proportional difference.

- The probability that two strings are equal can be inverse proportional to their edit distance.

- For other identifiers (social security numbers, etc.), the probability of equivalence can be a function that is robust to common misspellings. The checksum computations that are often defined for such identifiers can give a hint as to which misspellings are common.

- By default, the probability of two different literals being equal should be 0.

These functions can be designed depending on the application or on the specific ontologies. They can, e.g., take into account unit conversions (e.g., between Kelvin and Celcius). They could also perform datatype conversions (e.g., between *xsd:string* and *xsd:anyURI*) if necessary. The probabilities can then be plugged into Equation (13).

For our implementation, we chose a particularly simple equality function. We normalize numeric values by removing all data type or dimension information. Then we set the probability $\Pr(x \equiv y)$ to 1 if $x$ and $y$ are identical literals, to 0 otherwise. The goal of this work is to show that even with such a simple, domain-agnostic, similarity comparison between literals, our probabilistic model is able to align ontologies with high precision; obviously, precision could be raised even higher by implementing more elaborate literal similarity functions.

### 5.4 Parameters

Our implementation uses the following parameters:

1. The initial value $\theta$ for the equivalence of relations in the very first step of the algorithm. We show in the experiments that the choice of $\theta$ does not affect the results.

2. Similarity functions for literals. These are application-dependent. However, we show that even with the simple identity function, the algorithm performs well.

Therefore, we believe we can claim that our model has no dataset-dependent tuning parameters. Our algorithm can be (and in fact, was) run on all datasets without any dataset specific settings. This contrasts PARIS with other algorithms, which are often heavily dependent on parameters that have to be tuned for each particular application or dataset. Traditional schema alignment algorithms, for example, usually use heuristics on the names of classes and relations, whose tuning requires expertise (e.g., [22]). A major goal of the present work was to base the algorithm on probabilities and make it as independent as possible from the tuning of parameters. We are happy to report that this works beautifully.

In order to improve results further, one can use smarter similarity functions, as discussed in Section 5.3.

## 6. EXPERIMENTS

### 6.1 Setup

All experiments were run on a quad-core PC with 12 GB of RAM, running a 64bit version of Linux; all data was stored on a fast solid-state drive (SSD), with a peak random access bandwidth of approximately 50 MB/s (to be compared with a typical random access bandwidth of 1 MB/s for a magnetic hard drive).

Our experiments always compute relation, class, and instance equivalences between two given ontologies. Our algorithm was run until convergence (i.e., until less than 1 % of the entities changed their maximal assignment). We evaluate the instance equalities by comparing the computed final maximal assignment to a gold standard, using the standard metrics of precision, recall, and F-measure. For instances, we considered only the assignment with the maximal score. For relation assignments, we performed a manual evaluation. Since PARIS computes sub-relations, we evaluated the assignments in each direction. Class alignments were also evaluated manually. For all evaluations, we ignored the probability score that PARIS assigned, except when noted.

### 6.2 Benchmark Test

To be comparable to [18, 22, 25, 27], we report results on the benchmark provided by the 2010 edition of the ontology alignment evaluation initiative (OAEI) [11]. We ran experiments on two datasets, each of which consists of two ontologies.[10] For each dataset, the OAEI provides a gold standard list of instances of the first ontology that are equivalent to instances of the second ontology. The relations and classes are identical in the first and second ontology. To make the task more challenging for PARIS, we artificially renamed the relations and classes in the first ontology, so that the sets of instances, classes, and relations used in the first ontology are disjoint from the ones used in the second ontology.

For the person dataset, PARIS converged after just 2 iterations and 2 minutes. For the restaurants, PARIS took 3 iterations and 6 seconds. Table 1 shows our results.[11] We achieve near-perfect precision and recall, with the exception of recall in the second dataset. As reported in [18], all other approaches [22, 25, 27] remain below 80 % of F-measure for the second dataset, while only OBJECTCOREF [18] achieves an F-measure of 90 %. We achieve an F-measure of 91 %. We are very satisfied with this result, because unlike OBJECT-COREF, PARIS does not require any training data. It should be further noted that, unlike all other approaches, PARIS did not even know that the relations and classes were identical, but discovered the class and relation equivalences by herself in addition to the instance equivalences.

---

[10]We could not run on the third dataset, because it violates our assumption of non-equivalence within one ontology.

[11]Classes and relations accumulated for both directions. Values for OBJCOREF as reported in [18]. Precision and recall are not reported in [18]. OBJCOREF cannot match classes or relations.

| Dataset | System | Instances | | | | Classes | | | | Relations | | | |
|---|---|---|---|---|---|---|---|---|---|---|---|---|---|
| | | Gold | Prec | Rec | F | Gold | Prec | Rec | F | Gold | Prec | Rec | F |
| **Person** | PARIS | 500 | 100% | 100% | 100% | 4 | 100% | 100% | 100% | 20 | 100% | 100% | 100% |
| | OBJCOREF | | 100% | 100% | 100% | | - | - | - | | - | - | - |
| **Rest.** | PARIS | 112 | 95% | 88% | 91% | 4 | 100% | 100% | 100% | 12 | 100% | 66% | 88% |
| | OBJCOREF | | N/A | N/A | 90% | | - | - | - | | - | - | - |

Table 1: Results (precision, recall, F-measure) of instance, class, and relation alignment on OAEI datasets, compared with OBJECTCOREF [18]. The "Gold" columns indicate the number of equivalences in the gold standard.

### 6.3 Design Alternatives

To measure the influence of $\theta$ on our algorithm, we ran PARIS with $\theta = 0.001, 0.01, 0.05, 0.1, 0.2$ on the restaurant dataset. A larger $\theta$ causes larger probability scores in the first iteration. However, the sub-relationship scores turn out to be the same, no matter what value $\theta$ had. Therefore, the final probability scores are the same, independently of $\theta$. In a second experiment, we allowed the algorithm to take into account all probabilities from the previous iteration (and not just those of the maximal assignment). This changed the results only marginally (by one correctly matched entity), because the first iteration already has a very good precision. In a third experiment, we allowed the algorithm to take into account negative evidence (i.e., we used Equation (14) instead of Equation (13)). This made PARIS give up all matches between restaurants. The reason for this behavior turned out to be that most entities have slightly different attribute values (e.g., a phone number "213/467-1108" instead of "213-467-1108"). Therefore, we plugged in a different string equality measure. Our new measure normalizes two strings by removing all non-alphanumeric characters and lowercasing them. Then, the measure returns 1 if the strings are equal and 0 otherwise. This increased precision to 100%, but decreased recall to 70%. Our experience with YAGO and DBPEDIA (see next experiment) indicates that negative evidence can be helpful to distinguish entities of different types (movies and songs) that share one value (the title). However, in our settings, positive evidence proved sufficient.

### 6.4 Real-world Ontologies

| Ontology | #Instances | #Classes | #Relations |
|---|---|---|---|
| YAGO | 2,795,289 | 292,206 | 67 |
| DBPEDIA | 2,365,777 | 318 | 1,109 |
| IMDB | 4,842,323 | 15 | 24 |

Table 2: YAGO [30], DBPEDIA [2] and IMDB.

We wanted to test PARIS on real-world ontologies of a large scale, with a rich class and relation structure. At the same time, we wanted to restrict ourselves to cases where an error-free ground truth is available. Therefore, we first chose to align the YAGO [30] and DBPEDIA [2] ontologies, and then to align YAGO with an ontology built out of the IMDB[12].

YAGO *vs.* DBPEDIA. With several million instances, these are some of the largest ontologies available. Each of them has thousands of classes and at least dozens of relations. We took only the non-meta facts from YAGO, and only the manually established ontology from DBPEDIA, which yields the datasets described in Table 2. Both ontologies use Wikipedia identifiers for their instances, so that the ground truth for the instance matching can be computed trivially.[13] However, the statements about the instances differ in both ontologies, so that the matching is not trivial. The class structure and the relationships of YAGO and DBPEDIA were designed completely independently, making their alignment a challenging endeavor.

We ran PARIS for 4 iterations, until convergence. Table 3 shows the results per iteration. To compute recall, we counted the number of shared instances in DBPEDIA and YAGO. Since YAGO selects Wikipedia pages with many categories, and DBPEDIA selects pages with frequent infoboxes, the two resources share only 1.4 million entities. PARIS can map them with a precision of 90% and a recall of 73%. If only entities with more than 10 facts in DBpedia are considered, precision and recall jump to 97% and 85%, respectively.

PARIS assigns one class of one ontology to multiple classes in the taxonomy of the other ontology, taking into account the class inclusions. Some classes are assigned to multiple leaf-classes as well. For our evaluation, we excluded 19 high-level classes (such as *yagoGeoEntity*, *physicalThing*, etc.). Then, we randomly sampled from the remaining assignments and evaluated the precision manually. It turns out that the precision increases substantially with the probability score (see Figure 1). We report the numbers for a threshold of 0.4 in Table 3 (the number of evaluated sample assignments is 200 in both cases). The errors come from 3 sources: First, PARIS misclassifies a number of the instances, which worsens the precision of the class assignment. Second, there are small inconsistencies in the ontologies themselves (YAGO, e.g., has several people classified as *lumber*, because they work in the wood industry). Last, there may be biases in the instances that the ontologies talk about. For example, PARIS estimates that 12% of the people convicted of murder in Utah were soccer players. As the score increases, these assignments get sorted out. Evaluating whether a class is always assigned to its most specific counterpart would require exhaustive annotation of candidate inclusions. Therefore we only report the number of aligned classes and observe that even with high probability scores (see Figure 2 and Table 3) we find matches for a significant proportion of the classes of each ontology into the other.

---
[12]The Internet Movie Database, http://www.imdb.com
[13]We hid this knowledge from PARIS.

|  | Instances | | | | Classes | | | | | Relations | | | | |
|---|---|---|---|---|---|---|---|---|---|---|---|---|---|---|
| Change to prev. | Time | Prec | Rec | F | Time | YAGO ⊆ DBP | | DBP ⊆ YAGO | | Time | YAGO ⊆ DBP | | DBP ⊆ YAGO | |
|  |  |  |  |  |  | Num | Prec | Num | Prec |  | Num | Prec | Num | Prec |
| - | 4h04 | 86% | 69% | 77% | - | - | - | - | - | 19in | 30 | 93% | 134 | 90% |
| 12.4% | 5h06 | 89% | 73% | 80% | - | - | - | - | - | 21in | 32 | 100% | 144 | 92% |
| 1.1% | 5h00 | 90% | 73% | 81% | - | - | - | - | - | 21in | 33 | 100% | 149 | 92% |
| 0.3% | 5h26 | 90% | 73% | 81% | 2h14 | 137k | 94% | 149 | 84% | 24in | 33 | 100% | 151 | 92% |

Table 3: Results on matching YAGO and DBPEDIA over iterations 1–4

The relations are also evaluated manually in both directions. We consider only the maximally assigned relation, because the relations do not form a hierarchy in YAGO and DBPEDIA. In most cases one assignment dominates clearly. Table 4 shows some of the alignments. PARIS finds non-trivial alignments of more fine-grained relations to more coarse-grained ones, of inverses, of symmetric relations, and of relations with completely different names. There are a few plainly wrong alignments, but most errors come from semantic differences that do not show in practice (e.g., *burialPlace* is semantically different from *deathPlace*, so we count it as an error, even though in most cases the two will coincide). Recall is hard to estimate, because not all relations have a counterpart in the other ontology and some relations are poorly populated. We only note that we find alignments for half of YAGO's relations in DBPEDIA.

YAGO *vs.* IMDB. Next, we were interested in the performance of PARIS on ontologies that do not derive from the same source. For this purpose, we constructed an RDF ontology from the IMDB. IMDB is predestined for the matching, because it is huge and there is an existing gold standard: YAGO contains some mappings to IMDB movie identifiers, and we could construct such a mapping for many persons from Wikipedia infoboxes.

The content of the IMDB database is available for download as plain-text files.[14] The format of each file is *ad hoc* but we transformed the content of the database in a fairly straightforward manner into a collection of triples. For instance, the file `actors.list` lists for each actor $x$ the list of all movies $y$ that $x$ was cast in, which we transformed into facts $actedIn(x, y)$. Unfortunately, the plain-text database does not contain IMDB movie and person identifiers (those that we use for comparing to the gold standard). Consequently, we had to obtain these identifiers separately. For this purpose, and to avoid having to access each Web page of the IMDB Web site, which would require much too many Web server requests, we used the advanced search feature of IMDB[15] to obtain the list of all movies from a given year, or of all persons born in a certain year, together with their identifiers and everything needed to connect to the plain-text databases.

Since our IMDB ontology has only 24 relations, we manually created a gold standard for relations, aligning 15 of them to YAGO relations.

As Table 5 shows, PARIS took much longer for each iteration than in the previous experiment. The results are convincing, with an F-score of 92 % for the instances. This is a considerable improvement over a baseline approach that aligns entities by matching their *rdfs:label* properties (achieving 97 % precision and only 70 % recall, with an F-score of 82 %). Examining by hand the few remaining alignment errors revealed the following patterns:

- Some errors were caused by errors in YAGO, usually caused by incorrect references from Wikipedia pages to IMDB movies.

- PARIS sometimes aligned instances in YAGO with instances in IMDB that were not equivalent, but very closely related: for example, *King of the Royal Mounted* was aligned with *The Yukon Patrol*, a feature version of this TV series with the same cast and crew; *Out 1*, a 13-hour movie, was aligned with *Out 1: Spectre*, its shortened 4-hour variation.

- Some errors were caused by the very naïve string comparison approach, that fails to discover, e.g., that *Sugata Sanshirô* and *Sanshiro Sugata* refer to the same movie. It is very likely that using an improved string comparison technique would further increase precision and recall of PARIS.

PARIS could align 80 % of the relations of IMDB and YAGO, with a precision of 100 %. PARIS mapped half of the IMDB classes correctly to more general or equal YAGO classes (at threshold 0). It performs less well in the other direction. This is because YAGO contains mostly famous people, many of whom appeared in some movie or documentary on IMDB. Thus, PARIS believes that a class such as *People from Central Java* is a subclass of *actor*.

As illustrated here, alignment of instances and relations work very well in PARIS, whereas class alignment leaves still some room for improvement. Overall, the results are very satisfactory, as this constitutes, to the best of our knowledge, the first holistic alignment of instances, relations, and classes on some of the world's largest ontologies, without any prior knowledge, tuning, or training.

## 7. CONCLUSION

We have presented PARIS, an algorithm for the automated alignment of RDFS ontologies. Unlike most other approaches, PARIS computes alignments not only for instances, but also for classes and relations. It does not need training data and it does not require any parameter tuning. PARIS is based on a probabilistic framework that captures the interplay between schema alignment and instance matching in a natural way, thus providing a holistic solution to the ontology alignment problem. Experiments show that our approach works extremely well in practice.

---
[14]http://www.imdb.com/interfaces#plain
[15]http://akas.imdb.com/search/

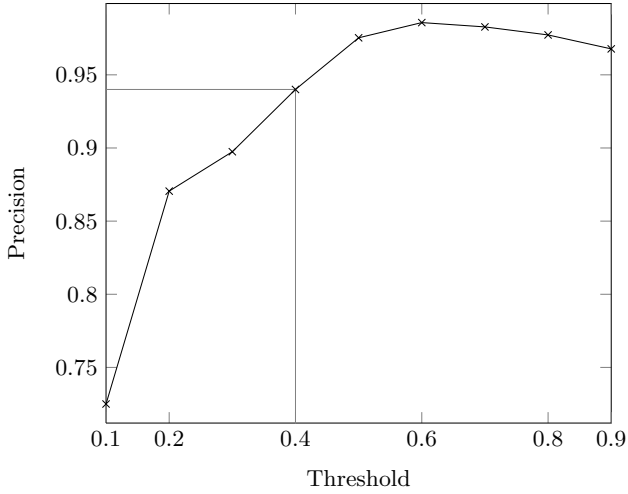

Figure 1: Precision of class alignment YAGO $\subseteq$ DBPEDIA as a function of the probability threshold.

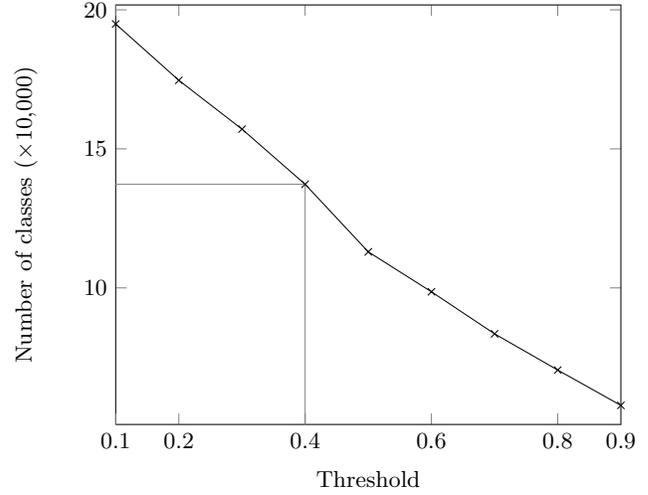

Figure 2: Number of YAGO classes that have at least one assignment in DBPEDIA with a score greater than the threshold.

| YAGO $\subseteq$ DBPEDIA | | | |
|---|---|---|---|
| y:actedIn | $\subseteq$ | dbp:starring$^{-1}$ | 0.95 |
| y:graduatedFrom | $\subseteq$ | dbp:almaMater | 0.93 |
| y:hasChild | $\subseteq$ | dbp:parent$^{-1}$ | 0.53 |
| y:hasChild | $\subseteq$ | dbp:child | 0.30 |
| y:isMarriedTo | $\subseteq$ | dbp:spouse$^{-1}$ | 0.56 |
| y:isMarriedTo | $\subseteq$ | dbp:spouse | 0.89 |
| y:isCitizenOf | $\subseteq$ | dbp:birthPlace | 0.25 |
| y:isCitizenOf | $\subseteq$ | dbp:nationality | 0.88 |
| y:created | $\subseteq$ | dbp:artist$^{-1}$ | 0.13 |
| y:created | $\subseteq$ | dbp:author$^{-1}$ | 0.17 |
| y:created | $\subseteq$ | dbp:writer$^{-1}$ | 0.30 |

| DBPEDIA $\subseteq$ YAGO | | | |
|---|---|---|---|
| dbp:birthName | $\subseteq$ | rdfs:label | 0.96 |
| dbp:placeOfBurial | $\subseteq$ | y:diedIn | 0.18 |
| dbp:headquarter | $\subseteq$ | y:isLocatedIn | 0.34 |
| dbp:largestSettlement | $\subseteq$ | y:isLocatedIn$^{-1}$ | 0.52 |
| dbp:notableStudent | $\subseteq$ | y:hasAdvisor$^{-1}$ | 0.10 |
| dbp:formerName | $\subseteq$ | rdfs:label | 0.73 |
| dbp:award | $\subseteq$ | y:hasWonPrize | 0.14 |
| dbp:majorShrine | $\subseteq$ | y:diedIn | 0.11 |
| dbp:slogan | $\subseteq$ | y:hasMotto | 0.49 |
| dbp:author | $\subseteq$ | y:created$^{-1}$ | 0.70 |
| dbp:composer | $\subseteq$ | y:created$^{-1}$ | 0.61 |

Table 4: Some relation alignments between YAGO and DBPEDIA with their scores

| | Instances | | | | Classes | | | | | Relations | | | | |
|---|---|---|---|---|---|---|---|---|---|---|---|---|---|---|
| Change to prev. | Time | Prec | Rec | F | Time | YAGO $\subseteq$ IMDB Num | Prec | IMDB $\subseteq$ YAGO Num | Prec | Time | YAGO $\subseteq$ IMDB Prec | Rec | IMDB $\subseteq$ YAGO Prec | Rec |
| - | 16h47 | 84% | 75% | 79% | - | - | - | - | - | 4min | 91% | 73% | 100% | 60% |
| 40.2% | 11h44 | 94% | 89% | 91% | - | - | - | - | - | 5min | 91% | 73% | 100% | 80% |
| 6.6% | 11h48 | 94% | 90% | 92% | - | - | - | - | - | 5min | 100% | 80% | 100% | 80% |
| 0.2% | 11h44 | 94% | 90% | 92% | 2h17 | 8 | 100% | 135k | 28% | 6min | 100% | 80% | 100% | 80% |

Table 5: Results on matching YAGO and IMDB over iterations 1–4

PARIS does not use any kind of heuristics on relation names, which allows aligning relations with completely different names. We conjecture that the name heuristics of more traditional schema-alignment techniques could be factored into the model.

Currently, PARIS cannot deal with structural heterogeneity. If one ontology models an event by a relation (such as *wonAward*), while the other one models it by an event entity (such as *winningEvent*, with relations *winner*, *award*, *year*), then PARIS will not be able to find matches. The same applies if one ontology is more fine-grained than the other one (specifying, e.g., cities as birth places instead of countries), or if one ontology treats cities as entities, while the other one refers to them by strings. For future work, we plan to address these types of challenges. We also plan to analyze under which conditions our equations are guaranteed to converge. It would also be interesting to apply PARIS to more than two ontologies. This would further increase the usefulness of PARIS for the dream of the Semantic Web.

## 8. ACKNOWLEDGMENTS

This work has been supported in part by the Advanced European Research Council grant Webdam on Foundations of Web Data Management, grant agreement 226513 (http://webdam.inria.fr/).

## APPENDIX
## A. GLOBAL FUNCTIONALITY

There are several design alternatives to define the *global functionality*:

1. We can count the number of statements and divide it by the number of pairs of statements with the same source:
$$fun(r) = \frac{\#x, y : r(x, y)}{\#x, y, y' : r(x, y) \wedge r(x, y')}$$
This measure is very volatile to single sources that have a large number of targets.

2. We can define functionality as the ratio of the number of first arguments to the number of second arguments:
$$fun(r) = \frac{\#x \quad \exists y : r(x, y)}{\#y \quad \exists x : r(x, y)}$$
This definition is treacherous: Assume that we have $n$ people and $n$ dishes, and the relationship *likesDish*$(x, y)$. Now, assume that all people like all dishes. Then *likesDish* should have a low functionality, because everybody likes $n$ dishes. But the above definition assigns a functionality of $fun(likesDish) = \frac{n}{n} = 1$.

3. We can average the local functionalities, as proposed in [17]:
$$fun(r) = \underset{x}{\operatorname{avg}} fun(r, x) = \underset{x}{\operatorname{avg}} \left( \frac{1}{\#y : r(x, y)} \right)$$
$$= \frac{1}{\#x \quad \exists y : r(x, y)} \sum_x \frac{1}{\#y : r(x, y)}.$$

However, the local functionalities are ratios, so that the arithmetic mean is less appropriate.

4. We can average the local functionalities not by the arithmetic mean, but by the harmonic mean instead
$$fun(r) = \underset{x}{\operatorname{HM}} fun(r, x) = \underset{x}{\operatorname{HM}} \left( \frac{1}{\#y : r(x, y)} \right)$$
$$= \frac{\#x \quad \exists y : r(x, y)}{\sum_x \#y : r(x, y)} = \frac{\#x \quad \exists y : r(x, y)}{\#x, y : r(x, y)}.$$

5. We may say that the global functionality is the number of first arguments per relationship instance:
$$fun(r) = \frac{\#x \quad \exists y : r(x, y)}{\#x, y : r(x, y)}$$
This notion is equivalent to the harmonic mean.

With these considerations in mind, we chose the harmonic mean for the definition of the global functionality.

## B. PROBABILISTIC MODELING

In Section 4, we presented a model of equality based on logical rules such as Equation (5), reproduced here:

$$\exists r, y : r(x, y) \wedge (\forall y' : r(x', y') \Rightarrow y \not\equiv y') \wedge fun(r) \text{ is high}$$
$$\implies x \not\equiv x$$

To transform these equations into probability assessments (Equation (6)), we assume mutual independence of all distinct elements of our models (instance equivalence, functionality, relationship inclusion, etc.). This assumption is of course not true in practice but it allows us to approximate efficiently the probability of the consequence of our alignment rules in a canonical manner. Independence allows us to use the following standard identities:

$$\Pr(A \wedge B) = \Pr(A) \times \Pr(B)$$
$$\Pr(A \vee B) = 1 - (1 - \Pr(A))(1 - \Pr(B))$$
$$\Pr(\forall x : \varphi(x)) = \prod_x \Pr(\varphi(x))$$
$$\Pr(\exists x : \varphi(x)) = 1 - \prod_x (1 - \Pr(\varphi(x)))$$
$$\mathbb{E}(\#x : \varphi(x)) = \sum_x \Pr(\varphi(x))$$

Then, a rule $\varphi \implies \psi$ can be translated into a probability assignment $\Pr(\psi) := \Pr(\varphi)$. $\varphi$ is recursively decomposed using the above identities. Following the example of Equation (5), we derive the value of $\Pr_2(x \equiv x')$ in Equation (6) as follows:

$$1 - \Pr\left(\exists r, y \ r(x, y) \wedge (\forall y' \ r(x', y') \Rightarrow y \not\equiv y') \wedge fun(r) \text{ is high}\right)$$
$$= \prod_{r,y} \left( 1 - \Pr(r(x, y)) \times \right.$$
$$\left. \prod_{y'} \left( 1 - \Pr(r(x', y') \wedge y \equiv y') \right) \times fun(r) \right)$$
$$= \prod_{r(x,y)} \left( 1 - fun(r) \prod_{r(x',y')} (1 - \Pr(y \equiv y')) \right)$$

since $r(x, y)$ and $r(x', y')$ are crisp, non-probabilistic facts.

Similarly, when we need to estimate a number such as "$\#x : \varphi(x)$", we compute $\mathbb{E}(\#x : \varphi(x))$ using the aforementioned identities.

The computed values reflect the probability that two entities are equal in the model, if the following conditions hold: (1) the literal equality functions of Section 5.3 return the probability of two values being equal, (2) a relation $r$ is either a function or not, and $fun(r)$ reflects the probability of this event, and (3) all probabilities are independent. Although these conditions are certainly not true to their full extent in practice, the equations still deliver useful approximations.

## C. EQUIVALENCE OF SETS

We compare two instances for equivalence by comparing every statement about the first instance with every statement about the second instance (if they have the same relation). This entails a quadratic number of comparisons. For example, if an actor $x$ acted in the movies $y_1$, $y_2$, $y_3$, and an actor $x'$ acted in the movies $y'_1$, $y'_2$, $y'_3$, then we will compare every statement $actedIn(x, y_i)$ with every statement $actedIn(x', y'_j)$. Alternatively, one could think of the target values as a set and of the relation as a function, as in $actedIn(x, \{y_1, y_2, y_3\})$ and $actedIn(x', \{y'_1, y'_2, y'_3\})$. Then, one would have to compare only two sets instead of a quadratic number of statements. However, all elements of one set are potentially equivalent to all elements of the other set. Thus, one would still need a quadratic number of comparisons.

One could generalize a set equivalence measure (such as the Jaccard index) to sets with probabilistic equivalences. However, one would still need to take into account the functionality of the relations: If two people share an e-mail address (high inverse functionality), they are almost certainly equivalent. By contrast, if two people share the city they live in, they are not necessarily equivalent. To unify two instances, it is sufficient that they share the value of one highly inverse functional relation. Conversely, if two people have a different birth date, they are certainly different. By contrast, if they like two different books, they could still be equivalent (and like both books). Our model takes this into account. Thus, our formulas can be seen as a comparison measure for sets with probabilistic equivalences, which takes into account the functionalities.

## D. REFERENCES